\documentclass[fleqn,usenatbib]{mnras}

\usepackage{newtxtext,newtxmath}

\usepackage[T1]{fontenc}

\DeclareRobustCommand{\VAN}[3]{#2}
\let\VANthebibliography\thebibliography
\def\thebibliography{\DeclareRobustCommand{\VAN}[3]{##3}\VANthebibliography}

\usepackage{graphicx}	
\usepackage{amsmath}	
\usepackage{subcaption}
\title[The mm Fundamental Plane]{A fundamental plane of black hole accretion at millimetre wavelengths}

\author[I. Ruffa et al.]{Ilaria Ruffa,$^{1,2}$\thanks{E-mail: ruffai@cardiff.ac.uk}
Timothy A. Davis,$^{1}$
Jacob S. Elford,$^{1}$\thanks{First-authorship is shared between Ruffa, Davis \& Elford}
Martin Bureau,$^{3}$
Michele Cappellari,$^{3}$
Jindra Gensior,$^{4}$
\newauthor
Daryl Haggard,$^{5}$
Satoru Iguchi,$^{6,7}$
Federico Lelli,$^{8}$
Fu-Heng Liang,$^{3}$
Lijie Liu,$^{9,10}$
Marc Sarzi,$^{11}$
\newauthor
Thomas G. Williams,$^{3}$
and Hengyue Zhang$^{3}$
\\
$^{1}$Cardiff Hub for Astrophysics Research \&\ Technology, School of Physics \&\ Astronomy, Cardiff University, Queens Buildings, The Parade, Cardiff, CF24 3AA, UK\\
$^{2}$INAF - Istituto di Radioastronomia, via P.\ Gobetti 101, 40129 Bologna, Italy\\
$^{3}$Sub-department of Astrophysics, Department of Physics, University of Oxford, Keble Road, Oxford, OX1 3RH, UK\\
$^{4}$Institute for Computational Science, University of Zurich, Winterthurerstrasse 190, Z{\"u}rich, 8057, Switzerland \\
$^{5}$Trottier Space Institute and Department of Physics, McGill University, 3600 rue University, Montreal, QC H3A 2T8, Canada\\
$^{6}$Department of Astronomical Science, SOKENDAI (The Graduate University of Advanced Studies), Mitaka, Tokyo 181-8588, Japan\\
$^{7}$National Astronomical Observatory of Japan, National Institutes of Natural Sciences, Mitaka, Tokyo 181-8588, Japan\\
$^{8}$INAF, Arcetri Astrophysical Observatory, Largo Enrico Fermi 5, I-50125, Florence, Italy\\
$^{9}$Cosmic Dawn Center (DAWN), Technical University of Denmark, DK2800 Kgs.\ Lyngby, Denmark\\
$^{10}$DTU-Space, Technical University of Denmark, Elektrovej 327, DK2800 Kgs.\ Lyngby, Denmark\\
$^{11}$Armagh Observatory and Planetarium, College Hill, Armagh BT61 9DG, UK
}

\date{Accepted XXX. Received YYY; in original form ZZZ}

\pubyear{2015}

\begin{document}
\label{firstpage}
\pagerange{\pageref{firstpage}--\pageref{lastpage}}
\maketitle

\begin{abstract}
We report the discovery of the ``mm fundamental plane of black-hole accretion'', which is a tight correlation between the nuclear 1 mm luminosity ($L_{\rm \nu, mm}$), the intrinsic $2$ -- $10$~keV X-ray luminosity ($L_{\rm X,2-10}$) and the supermassive black hole (SMBH) mass ($M_{\rm BH}$) with an intrinsic scatter ($\sigma_{\rm int}$) of $0.40$~dex. The plane is found for a sample of 48 nearby galaxies, most of which are low-luminosity active galactic nuclei (LLAGN). Combining these sources with a sample of high-luminosity (quasar-like) nearby AGN, we show that the plane still holds. We also find that $M_{\rm BH}$ correlates with $L_{\rm \nu, mm}$ at a highly significant level, although such correlation is less tight than the mm fundamental plane ($\sigma_{\rm int}=0.51$~dex). Crucially, we show that spectral energy distribution (SED) models for both advection-dominated accretion flows (ADAFs) and compact jets can explain the existence of these relations, which are not reproduced by the standard torus-thin accretion disc models usually associated to quasar-like AGN. The ADAF models reproduces the observed relations somewhat better than those for compact jets, although neither provides a perfect fit. Our findings thus suggest that radiatively-inefficient accretion processes such as those in ADAFs or compact (and thus possibly young) jets may play a key role in both low- and high-luminosity AGN. This mm fundamental plane also offers a new, rapid method to (indirectly) estimate SMBH masses.

\end{abstract}



\section{Introduction}
The many details of the processes regulating the connection between the growth of central super-massive black holes (SMBHs) and the evolution of their host galaxies (so-called "co-evolution”; e.g.\,\citealp{KormendyHo2013}) are still poorly understood \citep[e.g.][]{Donofrio21}. Understanding the physics of accretion onto SMBHs, determining if and how it changes in objects with different types of nuclear activity, as well as setting accurate constraints on fundamental SMBH properties such as its mass, are all crucial steps to get a comprehensive view of the SMBH-host galaxy interplay.

The so-called “fundamental plane of BH accretion” (heareafter FP) is an empirical correlation between the SMBH masses ($M_{\rm BH}$), 5~GHz radio ($L_{\rm 5 GHz}$) and 2 - 10~keV X-ray ($L_{\rm X,2-10}$) luminosities, which was initially reported by \citet{Merloni03} and \citet{Falcke04}. The origin of the FP is still debated, but it is widely believed to carry information on the physics of SMBH accretion (see e.g.\,\citealp[][]{Gultekin19}). However, the scatter around this correlation varies significantly depending on the sample and the method used to fit the plane, reaching values up to $0.88$~dex (e.g.\,\citealp[][]{Merloni03,Gultekin09,Plotkin12,Saikia15,Gultekin19}, and references therein). Furthermore, the nature of the radio emission in the FP is not yet well understood (potentially arising from compact jets or complex shock dynamics). All the above somehow limit the diagnostic power of the FP.

In this Letter, we report the discovery of a FP at millimetre wavelengths, namely the existence of a tight correlation between the nuclear (i.e.\,$\ll100$~pc) mm luminosities ($L_{\rm \nu, mm}$), $M_{\rm BH}$ and intrinsic $L_{\rm X,2-10}$, which we find to hold for both high- and low-luminosity AGN (within $z\lesssim0.05$). We also present the analysis of the physics underlying such correlation, and discuss how our results may have profound implications for our understanding of BH accretion in different AGN types.

\section{Primary sample and data}
\label{sec:methods-sample}
Our sample was primarily drawn from the mm-Wave Interferometric Survey of Dark Object Masses (WISDOM) project, which mainly exploits high-resolution Atacama Large Millimeter/submillimeter Array (ALMA) CO observations to dynamically estimate SMBH masses in a varied sample of galaxies \citep[e.g.][]{Davis17}. We included $31$ WISDOM galaxies (see Table~\ref{tab:datatable}) at $z\lesssim0.03$, spanning a range of AGN bolometric luminosities ($L_{\rm bol}=10^{41}$ -- $10^{46}$~erg~s$^{-1}$) and mostly (but not exclusively) having very low rates of accretion onto their central SMBHs ($\dot{M}\lesssim10^{-3}$~$\dot{M}_{\rm Edd}$; \citealp{Elford2023}). As such, most of these objects are classified as low-luminosity AGN (LLAGN; \citealp{Ho08}). To increase the statistics, we supplemented these 31 with a further 17 galaxies (see Table~\ref{tab:datatable}), selected from the literature to have dynamical SMBH masses, existing high-resolution ALMA $1$~mm and high-quality X-ray data. The majority of these are nearby ellipticals and span ranges of $L_{\rm bol}$ and $\dot{M}$ similar to those of the WISDOM sources. Hereafter, we refer to the 31 WISDOM plus 17 literature sources as the {\it primary sample}.

The 1 mm luminosities of the primary sample sources were derived from high-angular-resolution ALMA Band 6 continuum observations, taken between 2013 and 2021 as part of a large number of projects. 
All data were reduced using the Common Astronomy Software Applications ({\sc casa}) pipeline \citep{McMullin07}, adopting a version appropriate for each dataset and a standard calibration strategy. 
For more details on the data reduction see \cite{Davis22}.

For each dataset, continuum images were produced by combining the continuum spectral windows (SPWs) and the line-free channels of the line SPW (when included)  using the \textsc{CASA} task \textsc{tclean} in multi-frequency synthesis (MFS) mode. The resulting continuum maps have synthesised beams ranging from $0.\!\!^{\prime\prime}042$ to $0.\!\!^{\prime\prime}723$, corresponding to $6$ -- $330$~pc (average spatial resolution $\approx25$~pc). For each source, we measured the continuum flux density $f_{\rm mm}$ from the innermost synthesised beam, coincident with the galaxy core. The mm luminosities were then estimated as $L_{\nu,{\rm mm}}=4\pi D_{\rm L}^{2} f_{\rm mm} \nu_{\rm obs}$, where $D_{\rm L}$ is the luminosity distance and $\nu_{\rm obs}$ the observed frequency (between 231 and 239 GHz). As all the data were obtained with long-baseline configurations, extended dust emission is resolved out. We indeed typically detect only a point-like source at each galaxy centre, arising from unresolved core emission. In 3 (out of 48) galaxies, the emission is slightly resolved, making our measurements more uncertain. Removing these 3 objects, however, does not affect our results in any way.
The obtained $L_{\nu,{\rm mm}}$ are listed in Table~\ref{tab:datatable}.

The intrinsic (absorption-corrected) $2$--$10$~keV luminosities ($L_{\rm X,2-10}$) of the primary sample sources were retrieved from the literature, as detailed in \citet[][]{Elford2023}. In short, eight of these galaxies have no X-ray data available and were thus not considered in the parts of the analysis where $L_{\rm X,2-10}$ was required. For the vast majority of the objects with X-ray data (33/40), the adopted $L_{\rm X,2-10}$ was derived from {\it Chandra} observations, including only emission from the unresolved AGN core. For most of the {\it Chandra}-observed objects (26/33), accurate (intrinsic) nuclear $L_{\rm X,2-10}$ were retrieved from the catalogue of \citet[][see also Table~\ref{tab:datatable}]{Bi20}. 

Dynamically-determined SMBH masses (from stellar, ionised gas, molecular gas and/or maser kinematics) are available for a total of 31 primary sample sources (see Table~\ref{tab:datatable}). For the remaining 17 galaxies, we estimated $M_{\rm BH}$ using the $M_{\rm BH}$ --  $\sigma_{\star}$ relation of \cite{Vandenbosch16}, where $\sigma_{\star}$ is the stellar velocity dispersion within one effective radius. This was retrieved from the compilations of \citet{Vandenbosch16} and \citet{Cappellari2013} when available, from the HyperLeda database otherwise (\url{http://leda.univ-lyon1.fr}). Crucially, although constructed based on data availability only (and thus not meant to be complete in any statistical sense), our primary sample spans four orders of magnitude in SMBH mass.


\begin{figure*}
\centering
\includegraphics[clip=true, trim={22 22 17 17}, scale=0.38]{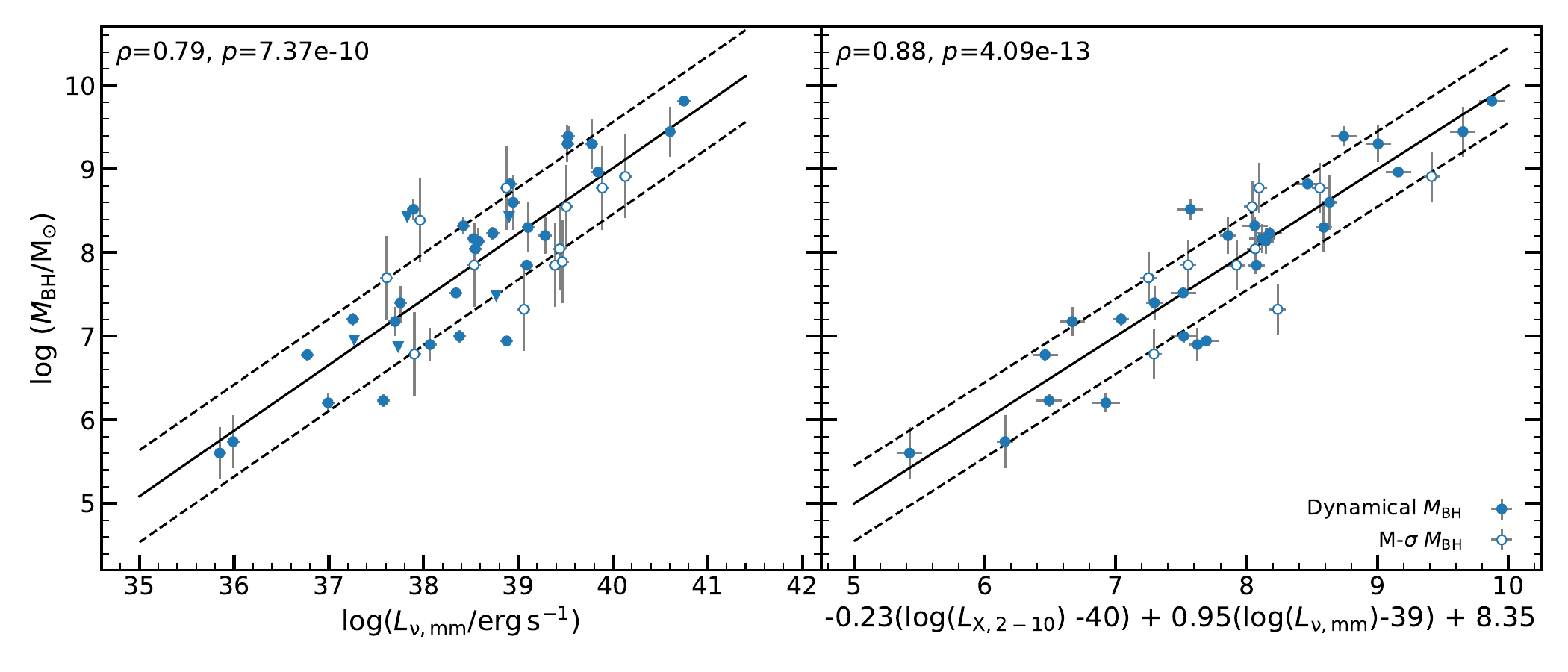}
\caption[]{Correlation between $M_{\rm BH}$ and $L_{\rm \nu, mm}$ (left panel) and edge-on view of the $M_{\rm BH}$-$L_{\rm X,2-10}$-$L_{\rm \nu, mm}$ correlation (right panel) for the primary sample galaxies. In both panels, filled blue circles show sources with dynamical $M_{\rm BH}$ measurements, open circles sources with $M_{\rm BH}$ from the $M_{\rm BH}$ -- $\sigma_{\star}$ relation of \cite{Vandenbosch16}. Error bars are plotted for all points but some are smaller than the symbol used. The best-fitting power-laws (Section~\ref{sec:methods-fitting}) are overlaid as a black solid lines, the observed scatter as black dashed lines. The correlation coefficients $\rho$ and $p$-values of the performed Spearman rank analysis are reported in the top-left corner of each panel.}\label{fig:fundamental_plane}
\end{figure*}

\vspace{-0.4cm}
\section{The mm fundamental plane}
\label{sec:methods-fitting}
As illustrated in Fig.\,\ref{fig:fundamental_plane} (left panel), the SMBH masses of our primary sample galaxies strongly correlate with their $L_{\rm \nu, mm}$. A power law was fitted to the observed trend, using the \textsc{lts\_linefit} routine \citep{Cappellari2013}. This combines the least-trimmed-squares (LTS) robust regression technique \citep{Rousseeuw84} with a least-squares fitting algorithm, and allows for intrinsic scatter and uncertainties in all coordinates. The resulting best-fitting power law is: 
\begin{equation}
    \log_{10}\left(\frac{M_{\rm BH}}{M_{\odot}}\right)=(0.79\pm0.08)\left[\log_{10}\left(\frac{L_{\nu, \rm mm}}{\mathrm{erg\,s}^{-1}}\right)-39\right] + (8.2\pm0.1)\,,
\end{equation}
with an observed scatter ($\sigma_{\rm obs}$) of $0.55$~dex and an estimated intrinsic scatter ($\sigma_{\rm int}$) of $0.51\pm0.08$~dex. When including $L_{\rm X,2-10}$, we discover the existence of a tighter correlation (Fig.\,\ref{fig:fundamental_plane}, right panel). In this case, we used the \textsc{lts\_planefit} routine \citep{Cappellari2013} to find the best-fitting plane in the ($\log M_{\rm BH}$, $\log L_{\rm X,2-10}$, $\log L_{\nu, \rm mm}$) space:

\begin{eqnarray}
    \nonumber \log_{10}\left(\frac{M_{\rm BH}}{M_{\odot}}\right) =  (-0.23\pm0.05)\left[\log_{10}\left(\frac{L_{\rm X,2-10}}{\mathrm{erg\,s}^{-1}}\right)-40\right] \\+   (0.95\pm0.07)\left[\log_{10}\left(\frac{L_{\nu,\rm mm}}{\mathrm{erg\,s}^{-1}}\right)-39\right] + (8.35\pm0.08)\,,
\end{eqnarray}
with $\sigma_{\rm obs}=0.45$~dex and $\sigma_{\rm int}=0.40\pm0.07$~dex. We verified that this multi-variate plane fit provides a significantly better predictor for $M_{\rm BH}$ than the simple line fit, having a $\Delta_{\rm BIC}$ $>>$10 (where $\Delta_{\rm BIC}$ is the difference in the Bayesian information criterion between the line and plane fits). For both correlations, we also performed Spearman rank analyses to quantify their statistical significance, and show the resulting correlation coefficients in the top-left corner of each panel of Fig.\,\ref{fig:fundamental_plane}. Since the nuclear mm and X-ray emission from AGN is known to be time variable (typically by a factor of $2$ -- $3$ over year timescales; \citealp{Prieto2016,Fernandez-Ontiveros2019,Behar20}), variability likely dominates the observed scatters (and thus the underlying correlations may be tighter). By analogy with the previous FP, we dub the correlation in the right panel of Fig.\,\ref{fig:fundamental_plane} as the ``mm fundamental plane of BH accretion" (hereafter mmFP).

We note that the error budget of the derived $L_{\nu,{\rm mm}}$ is dominated by the ALMA flux calibration uncertainties ($\approx$10\% for Band 6 data). These are, however, much smaller than the estimated intrinsic scatters of the correlations in Fig.\,\ref{fig:fundamental_plane} (see above), and thus have a negligible impact on our results. We also note that the $L_{\rm X,2-10}$ of the five X-ray observed sources without available {\it Chandra} data could be slightly overestimated, due to contamination from diffuse hot gas in the galactic and circum-galactic medium (CGM; although this mainly emits in the $0.3$ -- $2$~keV range) and/or X-ray binaries. While we verified that any such contamination should be minimal (based on the scaling laws of \citealp{Grimm03}, \citealp{Kim04} and \citealp{Boroson11}), we cannot rule it out entirely. 
In any case, removing these five sources does not make any relevant change in the best-fitting parameters of the mmFP. The same applies when removing the sources without a robust, dynamical $M_{\rm BH}$ estimate (the best-fitting line and planes are identical, within their respective errors, and the observed scatters become only slightly smaller).


\subsection{BASS galaxies}\label{sec:methods-bass}
\begin{figure}
\centering
\includegraphics[clip=true, trim={22 12 30 30}, scale=0.33]{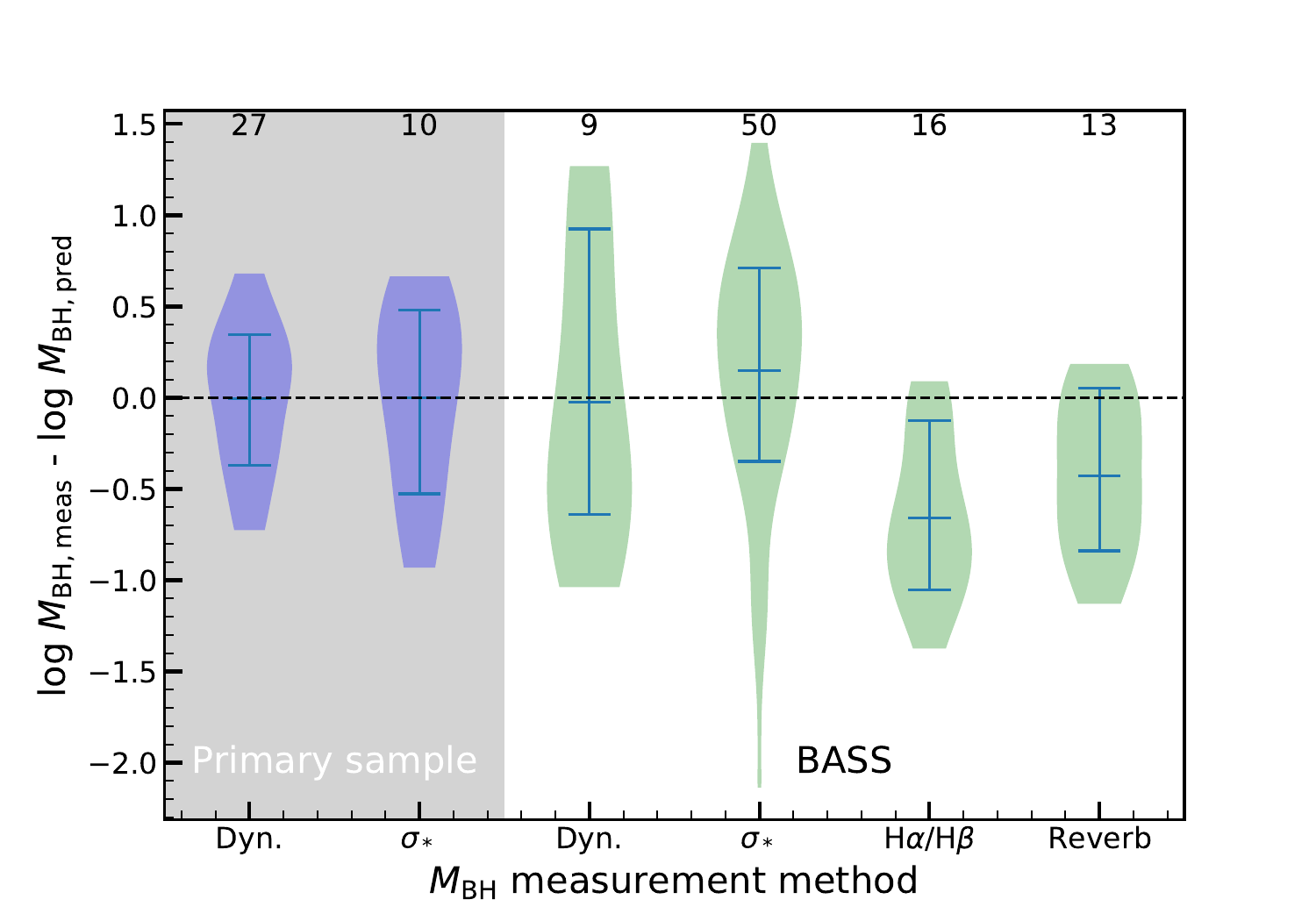}
\caption[]{Residuals of our primary sample (grey shaded region) and BASS sources (white region) from the best-fitting mmFP, plotted as a function of the SMBH mass measurement method. ``Dyn'' refers to dynamical mass measurements, ``$\sigma_{\star}$'' to estimates from the $M_{\rm BH}$ --  $\sigma_{\star}$ relation of \cite{Vandenbosch16}, ``H$\alpha$/H$\beta$'' to the broad-line method and ``Reverb'' to reverberation mapping. Each set of data points is represented by a violin describing the underlying distribution. The number of sources in the mmFP whose $M_{\rm BH}$ has been estimated using that particular method is indicated above each violin. In each case, the blue horizontal lines denote the $18^{\rm th}$, $50^{\rm th}$ and $85^{\rm th}$ percentiles of the distribution.}
\label{fig:bass_residuals}
\end{figure}

Although the majority of the primary sample galaxies are LLAGN, a handful are more luminous systems (see \citealp{Elford2023}), which still follow the mmFP. To investigate whether this result holds more generally, we built a comparison sample from the {\it Swift}-BAT AGN Spectroscopic Survey (BASS), comprising AGN with median $z=0.05$, $L_{\rm bol}=10^{44}$~erg~s$^{-1}$ and $\dot{M} = 0.01-0.1$~$\dot{M}_{\rm Edd}$ \citep[][]{Koss17}. We included only the BASS sources for which both ALMA $1$~mm observations (with spatial resolutions similar to those of the primary sample) and nuclear intrinsic $L_{\rm X,2-10}$ were available ($88$ sources; \citealp{Kawamuro2022}). The SMBH masses of these objects were taken from the compilation of \cite{Koss2022}. The BASS galaxies are typically more distant than those in the primary sample, so their $M_{\rm BH}$ have been estimated with a variety of methods (see Fig.\,\ref{fig:bass_residuals}). 
However, most of the sources ($50/88$) have their $M_{\rm BH}$ from the $M_{\rm BH}$ --  $\sigma_{\star}$ relation of \cite{KormendyHo2013}. We re-calibrated these measurements using the $M_{\rm BH}$ --  $\sigma_{\star}$ relation of \cite{Vandenbosch16}, for consistency with the 17 primary sample sources without a dynamical SMBH mass estimate.

As illustrated in Fig.\,\ref{fig:ADAF_model_projection}, the BASS sources are in agreement with the best-fitting mmFP, albeit with a larger observed scatter. We performed a Spearman rank analysis to quantify the statistical significance of this relation for the BASS points alone, and verified that they do show a significant correlation ($p=0.002$), but with a correlation coefficient ($\rho$=0.32) smaller than that of the primary sample. Fig.\,\ref{fig:bass_residuals} suggests that the larger scatter in this population is (at least partly) driven by the $M_{\rm BH}$ uncertainties, as the position of a BASS galaxy with respect to the best-fitting mmFP depends on the method used to estimate its $M_{\rm BH}$. 
For instance, sources with $M_{\rm BH}$ from reverberation mapping or broad-line methods are located systematically below the best-fitting line, likely reflecting the different biases in place when using such techniques \citep[e.g.][]{Farrah2023}. 

\section{Physical Drivers}\label{sec:models}
The fact that the BASS galaxies are consistent with a relation mainly defined by LLAGN is surprising. 
To determine the underlying physics, we compared the observed nuclear mm and X-ray luminosities of both the primary sample and BASS sources to those extracted from mock nuclear SEDs arising from both ``classic'' and radiatively-inefficient (ADAF-like) accretion flows, and from compact radio jets.

\subsection{Torus model}
\label{sec:methods-torusmodel}
AGN in the BASS sample (and a few in the primary sample) have {\it estimated} accretion rates in the range $\dot{M} \sim 0.01-0.1$~$\dot{M}_{\rm Edd}$. According to the standard paradigm, in this type of systems the accretion should occur through the classic geometrically-thin and optically-thick accretion disc surrounded by a dusty torus \citep[e.g.][]{Heckman14}. In this scenario, both the mm and the $2$ -- $10$~keV emission arise from the accretion disc, reprocessed by dust in the torus in the mm and Compton-up scattered by the hot corona in X-rays. To check if this type of model can reproduce the observed mm and X-ray luminosities, we used the SKIRTOR library \citep{Stalevski2012,Stalevski2016}. 
The SED models were retrieved from the SKIRTOR webpage ({\url{https://sites.google.com/site/skirtorus/}), but their spectral coverage (from $300$~GHz to $1.24$~keV) is slightly shorter than that required for this work. We thus expanded the models to the full range of wavelengths probed here, treating the emission mechanisms self-consistently as prescribed in the original version of the code (i.e.\,using the same grey-body curve for millimetre emission, and a power-law in the X-ray regime; \citealp[][]{Yang2020}). We followed the prescriptions of \cite{Stalevski2012,Stalevski2016} to scale the models for different $L_{\rm bol}$, in the range $10^{7.5}$ -- $10^{12.5}$~L$_{\odot}$ (i.e.\,the range covered by the primary and BASS sources; the torus is expected to disappear at low accretion rates, but the resulting model predictions are nevertheless instructive). For each SKIRTOR SED model, we then extracted the predicted $1$~mm (specifically, the luminosity at 237.5~GHz, that is the median ALMA continuum frequency for both the primary and BASS sources) and intrinsic 2-10~keV luminosities, and compared them with the measured ones. The resulting predictions are shown in Fig.\,\ref{fig:mm_xray_with_grid_and_skirtor} as a hexagonally binned histogram (coloured by mean L$_{\rm bol}$). The luminosities extracted from the torus models reasonably reproduces the slope of the $L_{\rm X,2-10}$ -- $L_{\rm \nu, mm}$ relation of the BASS sources, but with an offset of about two orders of magnitude at a given L$_{\rm bol}$. On the other hand, to explain $L_{\rm \nu, mm}$ of the lower accretion rate galaxies, the mm luminosities in the SKIRTOR models would need to be at least four orders of magnitude larger at a given accretion rate (and thus X-ray luminosity).

\subsection{ADAF model}
\label{sec:methods-adafmodel}
To build model SEDs arising from radiatively-inefficient accretion flows around SMBHs, we used the ``LLAGN'' model of \citet[][itself a development of previous models by \citealt{NarayanYi1995} and \citealt{Mahadevan1997}]{Pesce21}. In typical LLAGN and some (low-accretion-rate) Seyferts, the classic accretion disc is either absent or truncated at some inner radius (the transition usually happens beyond a few tens of Schwarzschild radii), and replaced by a geometrically-thick two-temperature structure in which the ion temperature is greater than the electron temperature and the accretion occurs at rates well below the Eddington limit (i.e.\,$\ll0.01$~$\dot{M}_{\rm Edd}$; \citealp{Nara95,Ho08}). The electrons in such radiatively-inefficient flows (such as advection-dominated accretion flows; ADAFs) cool down via a combination of self-absorbed synchrotron, bremsstrahlung and inverse Compton radiation, which together give rise to the nuclear SED from the mm to the X-rays. The LLAGN model adopted here solves for the energy balance between the heating and cooling of the electrons in the flow. We generated a set of model SEDs for a grid of SMBH masses ($10^{6}$ -- $10^{10}$~M$_{\odot}$) and Eddington ratios ($10^{-7}$ -- $10^{-2}$), while all the other free parameters were kept at the defaults discussed in Appendix~A of \cite{Pesce21}. We then extracted the predicted $237.5$~GHz and $2-10$~keV luminosities, as described above. As illustrated by the shape of the model grid in Fig.\,\ref{fig:mm_xray_with_grid_and_skirtor}, the mm and X-ray luminosities of all the sources (and thus the observed correlations) are well explained if they arise from an ADAF-like accretion mechanism. The grid is almost aligned with the axes, thus predicting that the mm luminosity primarily depends on $M_{\rm BH}$, while $L_{\rm X,2-10}$ primarily traces the Eddington ratio. The tighter correlation obtained when including $L_{\rm X,2-10}$ can be explained by the fact that the slight tilt of the grid is then taken into account, especially at higher Eddington ratios. A 3D version of Fig.\,\ref{fig:mm_xray_with_grid_and_skirtor} (including all the primary sample and only the BASS sources with the most robust, dynamical $M_{\rm BH}$ measurements; see Sect.\,~\ref{sec:methods-bass} and Fig.\,~\ref{fig:bass_residuals}) is provided as supplementary online material. In Fig.\,\ref{fig:ADAF_model_projection}, we show the projection of the ADAF model grid onto the best-fitting mmFP. This latter seems to arise naturally from these models, as an (almost) edge-on view of the 3D ($M_{\rm BH}$, $L_{\rm X,2-10}$, $L_{\rm \nu, mm}$) relation. We note, however, that keeping the default model parameters from \cite{Pesce21}, the model well predicts the gradient of the mmFP, but is offset by a small amount (i.e.\,the model overpredicts $L_{\rm \nu, mm}$ at a given SMBH mass by $\approx0.5$~dex). Tweaking the model parameters to reduce the effective radiative efficiency easily removes this offset (e.g.\,by changing some combination of the effective viscosity, ratio of gas to magnetic pressure, fraction of viscous heating going directly to the electrons, outer radius of the ADAF, and/or power-law index of the mass accretion rate as a function of radius). However, as the correct values of these parameters is not well constrained, here we simply offset the model grid by a constant $0.5$~dex in SMBH mass to align it with the observed correlation. We stress that this scaling factor is significantly smaller than the one required for torus models to reproduce the observed trend (see Section~\ref{sec:methods-torusmodel}), and is well within the uncertainties for the adopted model parameters (see \citealp{Pesce21}). Future work exploring the parameters of ADAF-like models in sources with more extensive (sub-)mm coverage will allow to better understand the observed correlation, the small offset, and the physics of accretion onto these SMBHs. This is discussed further in Section~\ref{sec:discussion}.

\begin{figure}
\centering
\begin{subfigure}[t]{0.3\textheight}
\caption{}\label{fig:mm_xray_with_grid_and_skirtor}
\includegraphics[clip=true, trim={22 22 18 16}, scale=0.43]{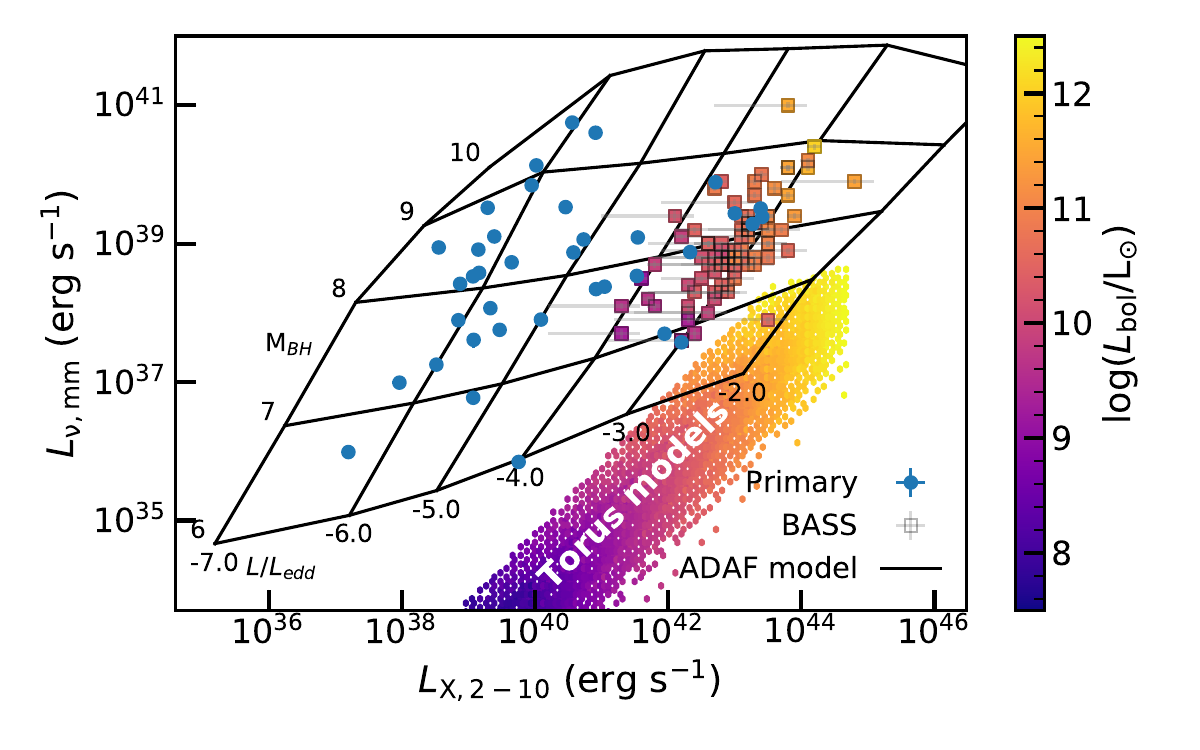}
\end{subfigure}
\vspace{0.2cm}
\medskip
\begin{subfigure}[t]{0.3\textheight}
\caption{}\label{fig:ADAF_model_projection}
\centering
\includegraphics[clip=true, trim={22 7 45 45}, scale=0.40]{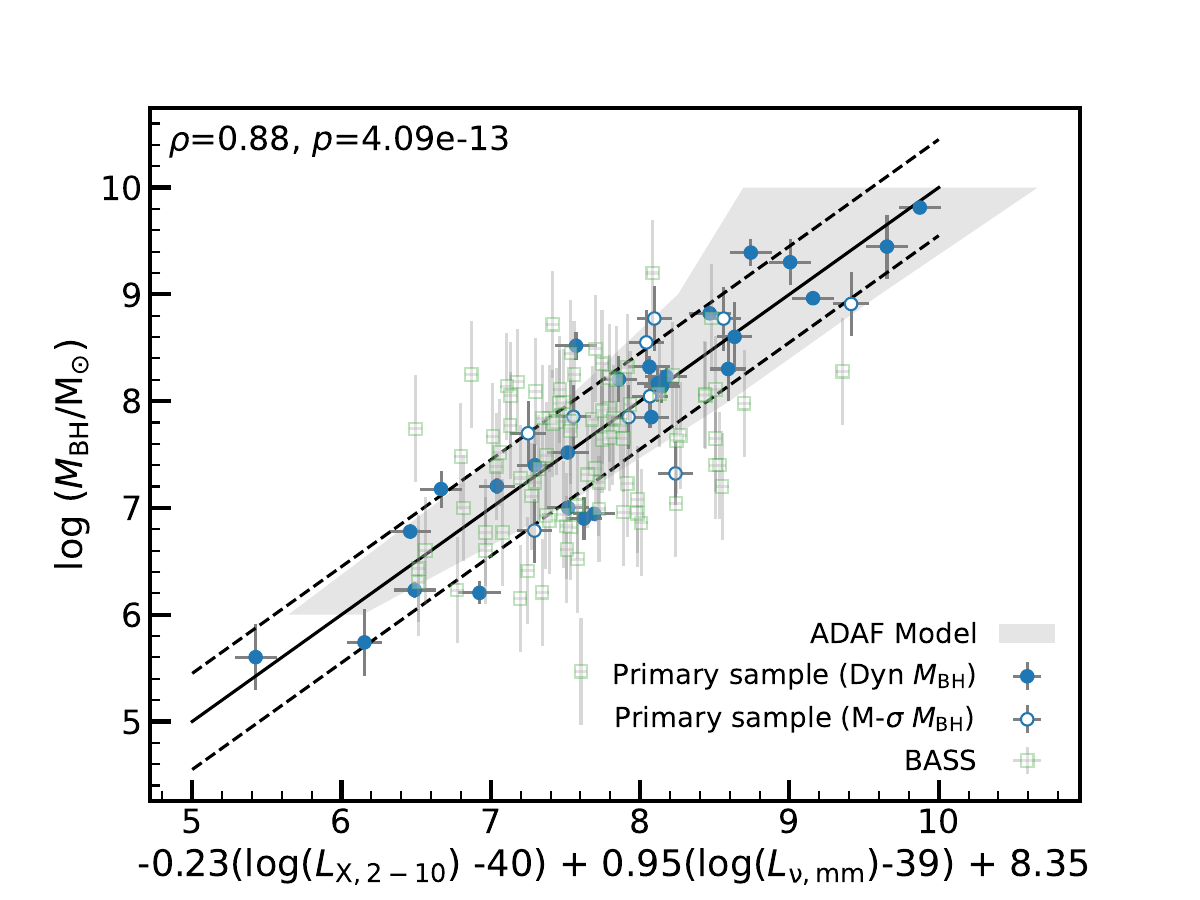}
\end{subfigure}
\vspace{-0.4cm}
\caption[]{{\bf Top}: Correlation between $L_{\rm X,2-10}$ and $L_{\rm \nu, mm}$ for the primary sample (blue circles) and the BASS (square symbols, coloured by their $L_{\rm bol}$) sources. Error bars are plotted for all points but some are smaller than the symbol used. The black grid illustrates the area covered by the ADAF model solutions as a function of $M_{\rm BH}$ and the Eddington ratio $L/L_{\rm Edd}$ (Section~\ref{sec:methods-adafmodel}). The purple--yellow coloured bins indicate the region covered by the extrapolated SKIRTOR torus models (Section~\ref{sec:methods-torusmodel}), where each hexagonal bin is coloured by the mean $L_{\rm bol}$ of the sources within that bin. {\bf Bottom}: As the right panel of Figure~\ref{fig:fundamental_plane}, but with the BASS galaxies overlaid as green squares and the grid of ADAF models from Figure~\ref{fig:mm_xray_with_grid_and_skirtor} projected onto the plane as a grey shaded area (including a small offset for clarity; see Section~\ref{sec:methods-adafmodel}).}\label{fig:model_results}
\end{figure}


\subsection{Compact jet model}
\label{sec:methods-compactjet}
Unlike extended jets (where the synchrotron emission is optically-thin), compact radio jets have self-absorbed synchrotron spectra (similar to those from ADAFs) and have been argued to dominate the nuclear SEDs of LLAGN. In some cases, they are preferred over a pure ADAF solution, as this would be overly luminous at near-infrared and optical wavelengths \citep[e.g.][]{Fernandez-Ontiveros2023}. To determine if compact jets can explain the observed trends, we used the \textsc{Bhjet} model of \cite{Lucchini2022}. We fixed most of the model parameters to the values found for M81, a prototypical AGN with compact jets (Model B in Table 3 of \citealt{Lucchini2022}), and generated a grid of models varying the SMBH mass ($10^{6}$\,--\,$10^{10}$~M$_{\odot}$), jet power ($10^{-5.5}$\,--\,$10^{-0.5}$ L$_{\rm Edd}$) and jet inclination to the line-of-sight (2.5$^{\circ}$\,--\,90$^{\circ}$). We then extracted from the resulting model SEDs the predicted $L_{\rm \nu, mm}$ and $L_{\rm X,2-10}$ (as above), and compare them with the observed ones. In Fig.\,\ref{fig:jetmodel_results} we show the $L_{\rm X,2-10}$ -- $L_{\rm \nu, mm}$ relation with overlaid the resulting model grids for the extremes in jet inclination (2.5$^{\circ}$ and 90$^{\circ}$). Jets at intermediate inclinations lie between these two extremes (but evolve quickly towards the i=90$^{\circ}$ solution once the line-of-sight is no longer aligned along the jet cone). 
The model grids encompass the majority of the LLAGN and some BASS sources, but they have significant curvature in the 3D $M_{\rm BH}$-$L_{\rm X,2-10}$-$L_{\rm \nu, mm}$ space (as for Fig.\,\ref{fig:mm_xray_with_grid_and_skirtor}, a 3D version of Fig.\,\ref{fig:jetmodel_results} is provided in the online material). The correlations in Fig.\,\ref{fig:fundamental_plane} do not seem to occur naturally within this model (as projections of the higher-order surface onto the axes). 
The luminosities of high-accretion-rate AGN from the primary and the BASS samples are harder to explain with these models, and would require additional X-ray emitting components. This is perhaps unsurprising, as compact jet models are substantially more complex than ADAFs (see also Section~\ref{sec:discussion}).

\begin{figure}
\centering
\includegraphics[clip=true, trim={20 20 20 17}, scale=0.45]{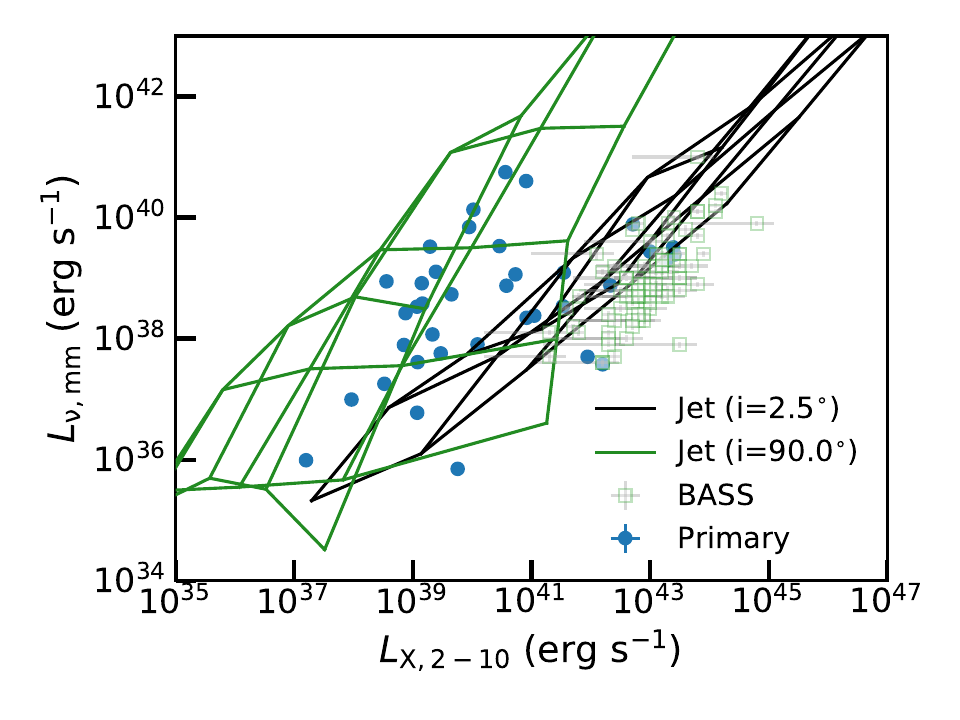}
\vspace{-0.3cm}
\caption[]{As Figure~\ref{fig:model_results}, but with slightly expanded axis ranges. In this case, the grids overlaid in black and green illustrate the areas covered by the compact jet model solutions as a function of $M_{\rm BH}$ and the jet power (Section~\ref{sec:methods-compactjet}), for jet inclinations of 2.5$^{\circ}$ and 90$^{\circ}$, respectively. Solutions with intermediate inclinations lie in-between these two extremes (see Section~\ref{sec:methods-compactjet}).}\label{fig:jetmodel_results}
\end{figure}


\subsection{Distance uncertainties}\label{sec:distance_check}
Both $M_{\rm BH}$ and luminosity measurements are systematically affected by the assumed galaxy distance $D$, with $M_{\rm BH}\propto D$ and $L\propto D^{2}$. Large distance errors can thus introduce large uncertainties on $M_{\rm BH}$ and $L$, and the difference in how these quantities scale with distance can give rise to spurious correlations. To test that this is not affecting our results, we performed a simple Monte Carlo simulation, drawing $M_{\rm BH}$ and the luminosities from independent Gaussian distributions that are truly uncorrelated, forcing a correlation to arise due to distance errors alone. The magnitude of the distance errors required to reproduce the Spearman rank correlation coefficients in Fig.\,\ref{fig:fundamental_plane} turned out to be $\geq$1.5 dex, much higher than that of our primary sample sources and - more in general - expected for real distance measurements. In addition, the slope of relations purely due to distance uncertainties would be substantially flatter than those observed (gradients of 0.5 for the $M_{\rm BH}$- $L_{ \rm \nu, mm}$ correlation and 0.25 for the mmFP, as opposed to the observed $\approx$0.8 and $\approx$1, respectively). 

To further check for any systematic distance bias, we carried out a simple quality-checking exercise for our primary sample. We restricted our analysis to only those sources with the most accurate (redshift-independent) distances (26/48), i.e.\,derived from surface brightness fluctuations, tip of the red giant branch methods, supernovae, Cepheids, masers, the planetary nebula luminosity function and the globular cluster luminosity function. This led us to obtain much tighter correlations, with an intrinsic scatter of $0.19$~dex for the $M_{\rm BH}-L_{\nu, \rm mm}$ relation and only $0.11$~dex for the mmFP. The corresponding Spearman rank coefficients are $\rho=0.84$ ($p=5.05\times10^{-7}$) and $\rho=0.93$ ($p=2.67\times10^{-9}$), respectively. We thus conclude that our results are not biased due to distance uncertainties. 

\vspace{-0.6cm}
\section{Discussion and conclusions}
\label{sec:discussion}
We report here the finding of tight $M_{\rm BH}-L_{\nu, \rm mm}$ and $M_{\rm BH}-L_{\rm X,2-10}-L_{\nu, \rm mm}$ correlations (Fig.\,\ref{fig:fundamental_plane}). We dub the latter the ``mm fundamental plane of BH accretion" and find it to hold for both low- (mostly WISDOM) and high- (mostly BASS) luminosity AGN. 
To understand the physics underlying the mmFP, we compared the observed trend with models predicting the emission from different nuclear mechanisms. We find that the results for both our sample and the BASS sources are best explained if their emission in the mm and X-rays primarily arises from an ADAF-like process, but cannot be explained by a classic torus model (see Fig.\,\ref{fig:model_results}). This suggests that some kind of radiatively-inefficient accretion process may play a role in both low- and high-luminosity AGN, at least in the range of luminosities and accretion rates probed by the sources included in this work. While torii are known to exist in many of these AGN, some regions around their SMBHs may be radiatively-inefficient. For instance, some accretion disc solutions allow discs to transition from ADAF-like to geometrically-thin (and vice versa at different radii), and ADAFs could also exist above and below classic accretion discs \citep{Mahadevan1997}. Although the exact conditions under which this applies are still to be investigated, it is clear that - if confirmed - our results will have profound implications for our understanding of BH accretion in many different types of AGN. 

We also explored the possibility that both the mm and X-ray emissions arise from compact (and thus probably young; \citealp{ODea20}) radio jets (Section~\ref{sec:methods-compactjet}). These have been argued to dominate the whole SEDs of LLAGN \citep[e.g.][]{Fernandez-Ontiveros2023} and have spectral properties similar to those of an ADAF at the wavelengths probed here. This is also consistent with one of the most popular scenarios for the origin of the radio FP of LLAGN, suggesting that the correlation arises from strongly sub-Eddington jet-dominated emission \citep[e.g.][]{Falcke04,Plotkin12}. The contribution of compact jets to the nuclear SEDs of radiatively-efficient, quasar-like AGN is instead still hotly debated \citep[e.g.][]{Fawcett20,Girdhar22}. Our results are marginally consistent with these scenarios, as we find that compact jet models can explain the correlations for most of the LLAGN, but additional X-ray emitting components are required in the higher-luminosity systems. 

In short, we demonstrated that ADAF-like models convincingly predict the mmFP. Compact jets are also a plausible explanation (at least for LLAGN), but the corresponding models do not reproduce the correlation as naturally as the ADAF-like ones. We caution, however, that the plasma physics underlying both the ADAF and compact jet models is not well constrained, and significant uncertainties are present in all the model parameters and how they interact. We thus conclude that, while ``classic'' torus models seem to be ruled out, either ADAF-like or compact jet emission have the potential to explain the observed trend. The presence of one (or more) of these mechanisms could even help explaining the increased far-infrared/sub-mm contribution attributed to AGN in some empirical SED models \citep[e.g.][]{Symeonidis2022}. The tight $L_{\rm X,2-10}$-$L_{\rm \nu, mm}$ correlation observed in Fig.\,\ref{fig:mm_xray_with_grid_and_skirtor} for the BASS sources is also consistent with the one reported by \citet{Ricci23} between the 100~GHz and 14-150~keV luminosities (see also \citealp{Behar18}), and our results add interesting clues onto its origin. Determining with certainty the relevant mechanism(s) giving rise to the observed correlations is beyond the scope of this work, but is crucial to further our understanding of the SMBH accretion/ejection processes in different AGN types.

Beyond carrying information on the nuclear physics, the correlations presented here provide new rapid methods to indirectly estimate the mass of SMBHs (or their accretion rates, if one has alternative, robust estimates of $M_{\rm BH}$ and $L_{\rm \nu, mm}$; see e.g.\,\citealp{Ricci23}). Although direct $M_{\rm BH}$ estimates can be obtained using a variety of techniques (e.g.\,stellar or gas kinematics, reverberation mapping), these typically require very time-consuming observational campaigns and currently have limited application beyond the local Universe. The ability to use nuclear mm and (optionally) X-ray luminosities allows $M_{\rm BH}$ estimates when dynamical measurements are not possible and/or the standard scaling relations are unusable (such as in dwarf or disturbed galaxies). It also allows $M_{\rm BH}$ predictions over a wider range of redshifts. At the high-mass end of the correlations, ALMA can allow us to constrain $M_{\rm BH}$ up to $z\approx0.3$ (and is limited more by angular resolution and frequency coverage than sensitivity). Proposed new interferometers (such as the next-generation Very Large Array, ngVLA) should be able to push this to $z=1$ and beyond. Large X-ray surveys that can provide complementary X-ray data are also ongoing (e.g.\,eROSITA), and next-generation satellites (such as the Advanced Telescope for High ENergy Astrophysics, {\it Athena}) will extend these to higher-$z$. We also note that the intrinsic scatter of the $M_{\rm BH}-L_{\nu, \rm mm}$ relation is comparable to that of the $M_{\rm BH}-\sigma_{\star}$ relation \citep[e.g.][]{Vandenbosch16}, and $\sigma_{\rm int}$ of the mmFP in Fig.\,\ref{fig:fundamental_plane} is comparable or even lower than that of its radio counterpart (depending on the sample used to fit the plane; see e.g.\,\citealp[][]{Merloni03,Falcke04,Plotkin12,Gultekin2019}). When restricting our analysis to only those primary sample sources with the most accurate (redshift-independent) distances, we obtain much tighter correlations (see Section~\ref{sec:distance_check}), with $\sigma_{\rm int}$ comparable to that of the tightest scaling relations in Astronomy (such as the Baryonic Tully-Fisher relation; e.g.\,\citealp{Lelli16,Lelli19}). 
This technique - if sufficiently verified - is thus well suited to constrain the details of SMBH-host galaxy co-evolution in regimes that have been difficult to access up to now \citep{Williams2023}.

\vspace{-0.6cm}
\section*{Acknowledgements}
IR and TAD acknowledge support from grant ST/S00033X/1 through the UK Science and Technology Facilities Council (STFC). MB and TGW were supported by STFC consolidated grant `Astrophysics at Oxford' ST/H002456/1 and
ST/K00106X/1. DH acknowledges support from the Canada Research Chairs (CRC) program, the NSERC Discovery Grant program, and the Canadian Tri-Agency New Frontiers in Research -- Explorations fund. This paper makes use of ALMA data. ALMA is a partnership of ESO (representing its member states), NSF (USA) and NINS (Japan), together with NRC (Canada), NSC and ASIAA (Taiwan), and KASI (Republic of Korea), in cooperation with the Republic of Chile. The Joint ALMA Observatory is operated by ESO, AUI/NRAO and NAOJ. We acknowledge also the usage of the HyperLeda database and the NASA/IPAC Extragalactic Database (NED).

\vspace{-0.6cm}
\section*{Data Availability}
The ALMA data used in this article are all available to download at the ALMA archive (\url{https://almascience.nrao.edu/asax/}). The calibrated data, final products and original plots generated for this research study will be shared upon reasonable request to the first author. The X-ray data have been retrieved from the catalogue of \citet{Bi20} or from the NASA/IPAC Extragalactic Database (NED; \url{https://ned.ipac.caltech.edu/}).


\bibliographystyle{mnras}
\bibliography{mybibliography} 




\appendix

\section{Data Table}

\begin{table*}
\caption{Full list and main parameters of the galaxies in the primary sample.}\label{tab:datatable}
\begin{tabular}{l l c c c c c c c c c}
\hline
Sample & Galaxy & D & log$M_{\rm BH}$ & $\Delta {\rm log}M_{\rm BH}$ & Method & log$L_{\rm \nu, mm}$ & ALMA project & log$L_{\rm X,2-10}$ & $\Delta$log$L_{\rm X,2-10}$ \\ 
 & & (Mpc) & (M$_{\odot}$) & (dex) & & (erg s$^{-1}$) & & (erg s$^{-1}$) & (dex) \\ 
 (1) & (2) & (3) & (4) & (5) & (6) & (7) & (8) & (9) & (10) \\ 
\hline
WISDOM & FRL49 & 85.7 & 8.20 & 0.2 & Dyn & 39.28 & a & 43.27 & 0.04 \\ 
 & FRL1146 & 136.7 & 7.85 & 0.30 & $\sigma_{\star}$ & $<38.76$ & a & 43.41 & 0.04 \\ 
 & MRK567 & 140.6 & 7.48 & 0.30 & $\sigma_{\star}$ & 39.39 & a & -- & -- \\ 
 & {\bf NGC0404} & 3.0 & 5.74 & 0.30 & Dyn & 35.99 & b & 37.20 & 0.04 \\ 
 & NGC0449 & 66.3 & 8.77 & 0.30 & $\sigma_{\star}$ & 38.87 & c & 40.58 & 0.04 \\ 
 & NGC0524 & 23.3 & 8.60 & 0.32 & Dyn & 38.94 & d & 38.55 & 0.04 \\ 
 & {\bf NGC0708} & 58.3 & 8.30 & 0.30 & Dyn & 39.10 & d & 39.39 & 0.04 \\ 
 & NGC1194 & 53.2 & 7.85 & 0.10 & Dyn & 39.09 & e & 41.54 & 0.04 \\ 
 & {\bf NGC1387} & 19.9 & 6.90 & 0.20 & Dyn & 38.07 & c & 39.33 & 0.04 \\ 
 & NGC1574 & 19.3 & 8.05 & 0.20 & Dyn & 38.55 & f & -- & -- \\ 
 & {\bf NGC2110} & 35.6 & 8.77 & 0.30 & $\sigma_{\star}$ & 39.89 & g & 42.71 & 0.04 \\ 
 & {\bf NGC3169} & 18.7 & 7.85 & 0.30 & $\sigma_{\star}$ & 38.53 & h & 41.53 & 0.04 \\ 
 & NGC3351 & 10.0 & 5.85 & 0.30 & $\sigma_{\star}$ & $<37.25$ & i & 38.74 & 0.04 \\ 
 & NGC3368 & 18.0 & 6.87 & 0.10 & Dyn & $<37.73$ & g & 39.30 & 0.05 \\ 
 & {\bf NGC3607} & 22.2 & 8.14 & 0.15 & Dyn & 38.58 & h & 39.16 & 0.04 \\ 
 & NGC4061 & 94.1 & 9.30 & 0.30 & Dyn & 39.78 & j & -- & -- \\ 
 & NGC4429 & 16.5 & 8.17 & 0.10 & Dyn & 38.08 & i & 39.12 & 0.10 \\ 
 & {\bf NGC4435} & 16.5 & 7.40 & 0.20 & Dyn & 37.76 & h & 39.47 & 0.04  \\ 
 & {\bf NGC4438} & 16.5 & 7.70 & 0.30 & $\sigma_{\star}$ & 37.61 & h & 39.08 & 0.04 \\ 
 & {\bf NGC4501} & 14.0 & 6.79 & 0.30 & $\sigma_{\star}$ & 37.90 & k & 40.09 & 0.04 \\ 
 & {\bf NGC4697} & 11.4 & 7.20 & 0.10 & Dyn & 37.25 & h & 38.52 & 0.04 \\ 
 & NGC4826 & 7.4 & 6.20 & 0.11 & Dyn & 36.99 & i & 37.96 & 0.11 \\ 
 & NGC5064 & 34.0 & 8.39 & 0.30 & $\sigma_{\star}$ & 37.96 & k & -- & -- \\ 
 & {\bf NGC5765b} & 114.0 & 7.32 & 0.30 & $\sigma_{\star}$ & 39.06 & e & 40.73 & 0.05 \\ 
 & NGC5806 & 21.4 & 6.95 & 0.30 & $\sigma_{\star}$ & $<37.26$ & c & -- & -- \\ 
 & {\bf NGC5995} & 107.5 & 8.55 & 0.30 & $\sigma_{\star}$ & 39.51 & a & 43.39 & 0.04 \\ 
 & NGC6753 & 42.0 & 8.42 & 0.30 & $\sigma_{\star}$ & $<37.82$ & k & -- & -- \\ 
 & NGC6958 & 30.6 & 7.89 & 0.30 & $\sigma_{\star}$ & 39.47 & k & -- & -- \\ 
 & {\bf NGC7052} & 51.6 & 8.91 & 0.30 & $\sigma_{\star}$ & 40.13 & j & 40.02 & 0.05 \\ 
 & {\bf NGC7172} & 33.9 & 8.05 & 0.30 & $\sigma_{\star}$ & 39.44 & l & 43.00 & 0.05 \\ 
 & PGC043387 & 95.8 & 8.42 & 0.30 & $\sigma_{\star}$ & $38.90$ & h & -- & -- \\ 
\hline
Literature & {\bf Circinus} & 4.2 & 6.23 & 0.08 & Dyn & 37.57 & m & 42.20 & 0.04 \\ 
 & {\bf IC1459} & 25.9 & 9.45 & 0.30 & Dyn & 40.60 & n & 40.91 & 0.10 \\ 
 & {\bf M87} & 16.5 & 9.81 & 0.04 & Dyn & 40.75 & o & 40.56 & 0.05 \\ 
 & {\bf NGC1316} & 19.9 & 8.23 & 0.08 & Dyn & 38.73 & p & 39.65 & 0.04 \\ 
 & {\bf NGC1332} & 22.3 & 8.82 & 0.04 & Dyn & 38.91 & q & 39.15 & 0.08 \\ 
 & {\bf NGC1380} & 17.1 & 8.17 & 0.17 & Dyn & 38.53 & r & 39.07 & 0.04 \\ 
 & NGC3227 & 17.0 & 7.18 & 0.17 & Dyn & 37.70 & s & 41.95 & 0.09 \\ 
 & NGC3245 & 21.5 & 8.32 & 0.10 & Dyn & 38.42 & t & 38.88 & 0.09 \\ 
 & NGC3393 & 53.6 & 7.52 & 0.03 & Dyn & 38.34 & e & 40.92 & 0.09 \\ 
 & NGC3489 & 12.0 & 6.78 & 0.07 & Dyn & 36.77 & u & 39.07 & 0.09 \\ 
 & NGC3504 & 13.6 & 7.00 & 0.07 & Dyn & 38.38 & v & 41.05 & 0.09 \\ 
 & {\bf NGC3585} & 20.6 & 8.52 & 0.13 & Dyn & 37.89 & t & 38.85 & 0.04 \\ 
 & {\bf NGC4374} & 18.5 & 8.96 & 0.05 & Dyn & 39.84 & w & 39.95 & 0.04 \\ 
 & {\bf NGC4388} & 19.8 & 6.94 & 0.01 & Dyn & 38.88 & x & 42.33 & 0.05 \\ 
  & {\bf NGC4395} & 4.4 & 5.60 & 0.31 & Dyn & 35.85 & y & 39.91 & 0.05 \\ 
 & {\bf NGC6861} & 27.3 & 9.30 & 0.22 & Dyn & 39.52 & r & 39.29 & 0.04 \\ 
& {\bf UGC2698} & 91.0 & 9.39 & 0.12 & Dyn & 39.53 & z & 40.46 & 0.09 \\ 
\hline
\end{tabular}
\parbox[t]{\textwidth}{\textit{Notes:} (1) Sub-sample. (2) Galaxy name. (3) Most accurate galaxy distance. (4), (5) and (6) SMBH mass, uncertainty, and measurement method. ``Dyn'' refers to dynamical measurements, ``$\sigma_{\star}$'' to estimates from the $M_{\rm BH}$ --  $\sigma_{\star}$ relation of \cite{Vandenbosch16}. (7) Millimetre luminosity. Errors are not reported because they are simply dominated by flux calibration uncertainties, which are $\approx$10\% for ALMA Band 6 (see Section 2 of the main paper). (8) Project code of the ALMA continuum observations, where a:\,2017.1.00904.S, b:\,2017.1.00572.S, c:\,2016.1.00437.S, d:\,2017.1.00391.S, e:\,2016.1.01553.S, f:\,2015.1.00419.S, g:\,2016.1.00839.S, h:\,2015.1.00598.S, i:\,2013.1.0049.S, j:\,2018.1.00397.S, k:\,2015.1.00466.S, l:\,2019.1.00363.S, m:\,2018.1.01321.S, n:\,2015.1.01572.S, o:\,2015.1.01352.S, p:\,2017.1.01140.S, q:\,2015.1.00896.S, r:\,2013.1.00229.S, s:\,2016.1.00254.S, t:\,2017.1.00301.S, u:\,2017.1.00766.S, v:\,2017.1.00964.S, w:\,2013.1.00828.S, x:\,2012.1.00139.S, y:\,2015.1.00597.S, z:\,2016.1.01010.S. (9) and (10) intrinsic (absorption-corrected) $2-10$~keV X-ray luminosity, and associated uncertainty. Galaxies highlighted in bold have their $L_{\rm X,2-10}$ taken from the catalogue of \citet{Bi20}.}
\end{table*}


\bsp	
\label{lastpage}
\end{document}